\documentclass[12pt]{amsart}
\usepackage[margin=1in]{geometry}

\usepackage{amsfonts,amsaddr}
\usepackage{bbm}

\def\bea#1\eea{\begin{align}#1\end{align}}
\def\bes #1\ees{\begin{split}#1\end{split}}
\newcommand{\be}{\begin{equation}}
\newcommand{\ee}{\end{equation}}

\def\GU{\operatorname{U{}}}

\title{Mirror Symmetry and Partition Functions}
\author{Travis Maxfield,}
\address{Department of Radiology, Duke University \\ Durham, NC 27708}
\email{travis.maxfield@duke.edu}
\author{David R.~Morrison,}
\address{Departments of Mathematics and Physics
\\ University of California, Santa Barbara\\ Santa Barbara, CA 93106}
\email{drm@math.ucsb.edu}
\author{M.~Ronen Plesser}
\address{Center for Geometry and Theoretical Physics, Box 90318 \\
Duke University \\ Durham, NC 27708-0318}
\email{plesser@cgtp.duke.edu}
\begin{document}

\begin{abstract}
Localization methods have produced explicit expressions for the sphere partition functions of (2,2) superconformal field theories. The mirror symmetry conjecture predicts an IR duality between pairs of Abelian gauged linear sigma models, a class of which describe families of Calabi-Yau manifolds realizable as complete intersections in toric varieties. 
We investigate this prediction for the sphere partition functions and find agreement between that of a model and its mirror up to the scheme-dependent ambiguities inherent in the definitions of these quantities.
\end{abstract}

\maketitle
\section{Introduction}
\label{s:intro}

A conformal field theory determines a space of deformations obtained through conformal perturbation theory by defining the deformed $n$-point correlation functions
\begin{align}
\langle \mathcal{O}_1(x_1)\ldots \mathcal{O}_n(x_n)\rangle(\lambda) = \langle \mathcal{O}_1(x_1)\ldots \mathcal{O}_n(x_n)e^{\sum_I\lambda^I\int\!d^2x\,\Phi_I(z)} \rangle
\end{align}
where $\mathcal{O}_i$ are any local operators and $\Phi_I$ are \emph{truly marginal} operators of dimension $(1,1)$. The integrals lead to divergences requiring  regularization but after this is performed the power series in $\lambda$ are believed to be convergent. The two-point functions of the truly marginal operators determine the \emph{Zamolodchikov metric}
\begin{align}\label{eq:Zmet}
g_{IJ} = |x-y|^4 \langle \Phi_I(x)\Phi_J(y)\rangle\ .
\end{align}
This structure was investigated in \cite{Belavin:1984vu,Zamolodchikov:1986gt,Kutasov:1988xb}.

When the theory enjoys $(2,2)$ superconformal symmetry we have additional structure, and the deformation spaces of such theories have been the subject of detailed study over the past three decades.  The $(2,2)$ superconformal algebra contains a $\GU(1)_R\times\GU(1)_L$ current algebra, the \emph{$R$-symmetry algebra}.  Truly marginal supersymmetry-preserving deformations are the top components of chiral or twisted chiral supermultiplets with charges $(1,\pm 1)$ under this symmetry. The deformation space with its metric (\ref{eq:Zmet}) factors locally as $\mathcal{M}_c\times\mathcal{M}_t$ (provided that the supersymmetry is not enhanced beyond $(2,2)$ \cite{arXiv:1611.03101}).

When the supersymmetry is not enhanced, the $R$-symmetry algebra produces a complex structure on each of these factors under which the (restricted) metric is K\"ahler.  Introducing complex local coordinates $\lambda_a$ on $\mathcal{M}_c$ and $\tilde{\lambda}_{\tilde a}$ on $\mathcal{M}_t$ we have
\begin{align}
g_{a\bar{b}} &= \partial_a\partial_{\bar{b}} K_c(\lambda,\bar{\lambda})\nonumber\\g_{\tilde{a}\bar{\tilde{b}}} &= \partial_{\tilde{a}} \partial_{\bar{\tilde{b}}} K_t(\tilde{\lambda},\bar{\tilde{\lambda}})\ .
\end{align}
The real function $K_c$ (resp.\ $K_t$) is a K\"ahler potential, defined locally in the patches of an open cover of $\mathcal{M}_c$ (resp.\ $\mathcal{M}_t$).  On the overlaps $U\cap U'$ these functions change by K\"ahler transformations, acting on $K_c$ for example as
\begin{align}\label{eq:Ktr}
K_c^U = K_c^{U'} - f^{UU'}_c(\lambda) - \bar{f}^{UU'}_c(\bar{\lambda})
\end{align}
for some local holomorphic function $f^{UU'}_c$.  A compact smooth K\"ahler manifold typically does not have a global K\"ahler potential, but the deformation spaces $\mathcal{M}_{t/c}$ are of course typically noncompact.

In 2012, supersymmetric localization methods were applied to $(2,2)$ gauged linear sigma models  \cite{Benini:2012ui,Doroud:2012xw} to compute the partition function on $S^2$. Up to a multiplicative factor associated to the conformal anomaly, this partition function is invariant under RG flow and so, for models flowing to $(2,2)$ superconformal IR fixed points, computes properties of the fixed points. Localization relies on the fact that the supersymmetry algebra on a round $S^2$ can be embedded in the $(2,2)$ algebra. This can be done in two ways, producing two partition functions $Z_{t/c}$ depending on $\tilde{\lambda}$, respectively $\lambda$.  The authors of \cite{Jockers:2012dk} conjectured that\footnote{All three original papers~\cite{Benini:2012ui,Doroud:2012xw,Jockers:2012dk} suppressed the dependence on the radius of the two-sphere.}
\begin{align}\label{eq:ZisK}
Z_{t/c} = \left(r/r_0\right)^{c/3}e^{-K_{t/c}}\ ,
\end{align}
where $r$ is the radius of $S^2$ and $r_0$ is a scheme-dependent constant.\footnote{In principle, we might obtain different scheme-dependent constants $r_{0,c}$ and $r_{0,t}$, but by adjusting the scheme we may assume (if we wish) that the constants are the same.} $K_{t/c}$ is a K\"ahler potential on the moduli space of the IR fixed point to which the UV model flows. Evidence for this conjecture was provided in~\cite{Jockers:2012dk,Gomis:2012wy,Gerchkovitz:2014gta}.

The localization calculation requires that some one-loop determinants
be regulated, amounting to a choice of renormalization scheme. The
Zamolodchikov metric and the complex structure are expected
to be scheme-independent properties of the superconformal field
theory.  This means the partition function, if \eqref{eq:ZisK} holds,
is determined up to multiplication by the square of a local, possibly
scheme-dependent, holomorphic function.  

This issue was addressed in \cite{Gomis:2015yaa} from a novel perspective. Working directly with the superconformal theory (unlike the localization methods which use a specific UV completion), this work considered the anomalous dependence of the partition function on Weyl transformations of the spacetime metric while promoting the parameters $\lambda$ (resp.\ $\tilde{\lambda}$) to chiral (resp.\ twisted chiral) multiplets of (2,2) supersymmetry. Assuming that conformal perturbation theory can be regularized preserving supersymmetry, these authors demonstrated that 
\begin{itemize}
\item the conjecture \eqref{eq:ZisK} holds.
\item the exponentials 
$e^{-f^{UU'}_c}$ of the
transition functions
in \eqref{eq:Ktr} form the transition functions of a holomorphic line bundle $\mathcal{L}_{K_c}$ on $\mathcal{M}_c$. This means the K\"ahler metric on $\mathcal{M}_c$ is \emph{Hodge}. Similar statements hold for $\mathcal{M}_t$, whose K\"ahler metric is also Hodge.
\end{itemize}
The results of \cite{Gomis:2015yaa} show, essentially, that the effective action for the $\lambda$ multiplets is determined by a holomorphic section of $\mathcal{L}_{K_c}$. The line bundle $\mathcal{L}_{K_c}$, if nontrivial, is an obstruction to the existence of a nowhere-vanishing globally defined partition function. It also means that the effective action for the chiral multiplets taking values in $\mathcal{M}_c$ is not globally defined. If the parameters were dynamical, a nontrivial $\mathcal{L}_{K_c}$ would indicate an inconsistency of the theory; since they are not, this becomes a characteristic property of the theory.\footnote{This is analogous to a sigma model anomaly for the non-dynamical scalars, as was pointed out in \cite{Tachikawa:2017aux}.} Again, similar statements hold for the twisted chiral multiplets and the line bundle 
$\mathcal{L}_{K_t}$.\footnote{The authors of \cite{Gomis:2015yaa} also
considered four-dimensional theories.  In that case, it was shown in
\cite{compact-curves}
that the corresponding line bundle is not trivial in general, and
consequently there is more to the anomaly story, uncovered in \cite{Tachikawa:2017aux}.}

The $(2,2)$ superconformal algebra possesses a $\mathbb{Z}_2$ automorphism, the \emph{mirror automorphism}, under which the deformation spaces $\mathcal{M}_c$ and $\mathcal{M}_t$ are exchanged. A \emph{mirror pair} of quantum field theories flow to infrared fixed points differing only by this automorphism. One class of mirror pairs is furnished by non-linear sigma models with Calabi-Yau target space, which possess $(2,2)$ supersymmetry. Mirror symmetry in this context is the nontrivial statement that if $X$ and $Y$ are a mirror pair of Calabi-Yau manifolds, the IR dynamics of the two sigma models is governed by the same SCFT, with the natural mapping induced by the mirror automorphism
\cite{Dixon,LVW,Aspinwall:1990xe,CLS,Greene:1990ud,Candelas:1990qd,cdgp}.

An alternative UV-free model flowing to a $(2,2)$ superconformal field
theory is an Abelian GLSM \cite{Witten:1993yc}.\footnote{There are
  some preliminary results
% which indicate that this phenomenon is also present 
for some non-Abelian GLSMs \cite{Gu:2018fpm} but for simplicity in this paper we shall restrict ourselves to the Abelian case.} For suitable choices of the parameters these can flow to the same IR fixed points as non-linear sigma models on Calabi-Yau $X$ given by complete intersection subspaces in toric varieties. For such Calabi-Yau manifolds, the mirror $Y$ is given by a conjecture of Batyrev and 
Borisov~\cite{Batyrev:1994hm,1993alg.geom.10001B},\footnote{Generally, only certain aspects of this duality, mostly topological, have been mathematically proven to date \cite{complete1,complete2}.}  which includes mirror duals constructed earlier by Greene and Plesser \cite{Greene:1990ud}. This mirror duality, translated to the data defining a GLSM 
in~\cite{Aspinwall:1993rj,Candelas:1994bu,Morrison:1995yh}, implies a corresponding infrared duality of the linear models. This duality can be tested with the localization results of~\cite{Benini:2012ui,Doroud:2012xw,Doroud:2013pka}, which 
enable the computation of same quantity, the partition functions $Z_{t/c}$, from two different UV descriptions. We perform that test here, and find agreement between $Z_t$ of the original theory and $Z_c$ of the mirror theory to within the scheme-dependent ambiguity in their definition.

Predictions of mirror symmetry for the sphere partition functions have been analyzed previously~\cite{Benini:2012ui,Gomis:2012wy,Doroud:2013pka,Benini:2016qnm}. However, the form of mirror symmetry that these authors tested and confirmed was that of Hori and Vafa~\cite{Hori:2000kt}. The relationship between this mirror symmetry and that of Batyrev and Borisov is unclear. Our calculation will shed some light on this relationship, but questions remain.

This note is structured as follows: in section~\ref{sect:GLSMs}, we
will review the structure of Abelian gauged linear sigma models and
the mirror map between mirror pairs. Additionally, we will review the
results of localization calculations for both partition functions
$Z_{t/c}$. In section~\ref{sect:mirrorpartitions}, we will demonstrate
the relationship between the partition functions of a model and its
dual. Appendix~\ref{sect:asystem} contains a demonstration that $Z_c$ 
of the mirror model satisfies a set of system of differential equations shown in~\cite{Halverson:2013eua} to be solved by $Z_t$, a further test of mirror symmetry. Appendix~\ref{sect:abelianduality} contains some technical comments on Abelian duality for chiral/twisted chiral multiplets on the sphere.

\section{Abelian Gauged Linear Sigma Models} \label{sect:GLSMs}

An Abelian GLSM is a $(2,2)$ gauge theory constructed with $n$ chiral multiplets $\Phi_i$ transforming effectively under an Abelian gauge group $G = \GU(1)^{n-d}\times\Gamma$ for some discrete Abelian group $\Gamma$. The continuous symmetry is gauged by $n-d$ vector multiplets $V_a$ with invariant field strengths $\Sigma_a$. The discrete gauging is implemented as an orbifold. The models of interest also include a superpotential interaction given by a holomorphic gauge invariant polynomial $W(\Phi)$. The action is written in superspace as
\begin{align}\label{eq:glsm}
\mathcal{L} &= \int\!d^4\theta\left(\sum_i\overline{\Phi}_i e^{2Q_i^a V_a} \Phi_i-\frac{1}{4e^2}\sum_a|\Sigma_a|^2\right) \\
&+ \mu \int\!d\theta^+d\theta^-\, W(\Phi)  + \frac{i}{2\sqrt 2}\int\!d\theta^+d\bar{\theta}^- \Sigma_a \tau^a + c.c.
\end{align}
Here $\tau^a = \frac{\theta^a}{2\pi} + i \xi^a$ is a complexified FI term.

A family of GLSMs is characterized by $G$ and the choice of the monomials appearing in $W$, and parameterized by the continuous parameters in \eqref{eq:glsm}. These are $\tau^a$ and the coefficients of terms in $W$.  In particular, the discrete group $\Gamma$ does not appear in the Lagrangian. 

A convenient way to 
specify 
these data is to introduce, following \cite{Candelas:1994bu,Morrison:1995yh}, an $\tilde{n}\times n$ matrix $P$ of nonnegative integers and a factorization $P=\tilde{T} T$ of this into two integer valued rank-$d$ matrices. The rows of $T$ can be used to construct a collection of Laurent monomials $\Phi^{I}:=\prod_{i=1}^n \Phi_i^{T_{I i}}$, and the group $G$ is defined to be the largest subgroup of $H=\GU(1)^n$ which leaves the monomials $\Phi^{I}$ invariant. The monomials $\Phi^{\tilde{\imath}} := \prod_{i = 1}^n \Phi_i^{P_{\tilde\imath i}}$ defined by the rows of $P$ are then $G$-invariant by construction, thanks to the relation $P_{{\tilde{\imath}} i}=\sum_I \tilde T_{{\tilde{\imath}} I}T_{I i}$. That is, in this language, the gauge charges $Q^a_i$ are a basis for the kernel of $P$. Since $P_{{\tilde{\imath}} i}\ge0$ by assumption, we may use these monomials to specify the family of interaction polynomials
\begin{equation}\label{eq:eW}
W(\Phi ):=\sum_{{\tilde{\imath}}=1}^{\tilde{n}} c_{\tilde{\imath}} \Phi^{\tilde \imath}= \sum_{{\tilde{\imath}}=1}^{\tilde{n}} c_{\tilde{\imath}} \prod_{i=1}^n \Phi_i^{P_{{\tilde{\imath}} i}}~,
\end{equation}
where $c_{\tilde{\imath}}\in \mathbb{C}^\ast$ parameterize the family.
Alternatively, if we are given $G$ and a family of polynomials $W$, it is not difficult to reconstruct the matrices $P$, $\tilde{T}$, and $T$.  (Actually, $\tilde{T}$ and $T$ are only well-defined up to $(\tilde{T},T)\mapsto (\tilde{T} L,L^{-1}T)$, with $L$ an invertible integer matrix.) Conditions on $\tilde{T}$ and $T$ (beyond their rank) ensuring that the generic model in the family is nonsingular were discussed in~\cite{Aspinwall:2015zia}.

The model will flow at energies much smaller than $e$ or $\mu$ to a conformal field theory if the gauge action is such that $\prod_i \Phi_i$ is invariant (implying $\sum_iQ^a_i = 0$), and if there exists an assignment $\rho_i$ of rational $R$-charges such that $\rho(W) = 2$.\footnote{We are referring to the vector $R$-charges.} The latter is equivalent to $\sum_i P_{\tilde \imath i} \rho_i = 2, \forall \tilde \imath$. The central charge of the resulting theory obeys
\begin{align}\label{eq:c3}
{c\over 3} = \sum_{i=1}^n (1-\rho_i) - (n-d)~.
\end{align}
The $R$-charge assignment $\rho$ will play a role in our discussion.
In the localization computation of
\cite{Benini:2012ui,Doroud:2012xw,Jockers:2012dk} this determines the
coupling of the GLSM to the curvature of $S^2$ via an embedding of the
of the $(2,2)$ rigid supersymmetry algebra on the sphere into the
$(2,2)$ superconformal algebra. It is clear that $\rho$ is only
defined up to mixing with the gauge symmetry. This mixing has no
effect on the IR theory
or on the UV theory in the plane,
but it does change the UV completion on $S^2$ and thus the renormalization scheme in which the sphere partition function is calculated. The consequences of this observation will factor into the identification of the partition functions of the mirror models.

One advantage of the GLSM is that (some of) the parameters determining
the low-energy theory are explicitly clear.  The manifest chiral
deformations are parameterized by the coefficients
$c_{\tilde{\imath}}$ of $W$.  These can fail to provide global coordinates
on $\mathcal{M}_c$ in three ways:
\begin{itemize}
\item In general, $c_{\tilde{\imath}}$ parameterize the subspace of $\mathcal{M}_c$ representing theories obtainable as low-energy limits of GLSMs of the form \eqref{eq:glsm}, the \emph{toric subspace} of $\mathcal{M}_c$.   
\item The $c_{\tilde{\imath}}$ can overparameterize the toric subspace. Values of these related by transformations of the form
\begin{align}\label{eq:rescale}
c_{\tilde{\imath}}\to \lambda_i^{P_{{\tilde{\imath}} i}} c_{\tilde{\imath}}
\end{align}
describe the same models up to the irrelevant field redefinition
\begin{align}\label{eq:cstar}
\Phi_i\to \lambda^{-1}_i\Phi_i, \qquad \lambda\in (\mathbb{C}^\ast)^n\ .
\end{align}
Invariant coordinates are provided by 
\begin{align}\label{eq:eqt}
{\tilde q}_{\tilde a} = \prod_{\tilde{\imath}} c_{\tilde{\imath}}^{\tilde{Q}_{\tilde{\imath}}^{\tilde a}}\in \left(\mathbb{C}^\ast\right)^{\tilde{n}-d}\ ,
\end{align}
where $\tilde{Q}_{\tilde{\imath}}^{\tilde a}$, ${\tilde a} = 1,\ldots, \tilde{n}-d$ are a basis for the cokernel of $P$. In general there may be additional identifications
on the space of $c_{\tilde{\imath}}$.  In the cases of interest here these can be ``fixed'' by setting some of the coefficients to zero \cite{Aspinwall:1993rj}, maintaining
the rank of $P$.
\item The toric subspace of $\mathcal{M}_c$ includes not just $\left(\mathbb{C}^\ast \right)^{\tilde n - d}$ but a partial compactification of this space which includes, e.g., Gepner models. Additionally, there is a complex codimension-one subvariety $\Delta_c\in \mathbb{C}^{\tilde{n}-d}$ (and a corresponding compactification) for which the data do not determine a superconformal fixed point. This might contain some of the coordinate hyperplanes.
\end{itemize}

Similarly, the exponentiated complexified FI terms
\begin{align}
q_a =e^{2\pi i \tau^a}\in \left(\mathbb{C}^\ast\right)^{n-d}
\end{align}
provide coordinates on $\mathcal{M}_t$; more precisely they are holomorphic coordinates in an open neighborhood on the subspace of $\mathcal{M}_t$ describing theories arising as IR limits of GLSMs (the \emph{toric subspace}).  In general some of these may be redundant parameters. A complex codimension one subvariety $\Delta_t\in \left(\mathbb{C}^\ast\right)^{n-d}$ of these correspond to singular models and do not flow to superconformal fixed points.

The mirror map for GLSMs takes a particularly simple form, anticipated
in the notation above \cite{Aspinwall:1993rj,Candelas:1994bu,Morrison:1995yh}.  A
model constructed with $\tilde{n}$ chiral multiplets and $\tilde{n}-d$
vector multiplets, with the gauge representation determined by
$\tilde{Q}$ will flow to the 
same superconformal fixed point with operators mapped by the mirror automorphism
if the parameters $(\tilde q_{\tilde a}, \tilde c_i)$ are chosen such that
\be\label{eq:mondivmap}
\tilde q_{\tilde a} = \prod_{\tilde \imath=1}^{\tilde n} c_{\tilde \imath}^{\tilde Q^{\tilde a}_{\tilde \imath}}, \qquad q_a = \prod_{i=1}^n \tilde c_i^{Q^a_i}.
\ee
If there is a nontrivial factorization we also exchange $\tilde T$ with $T^T$. In other words, the dual model exchanges $P$ for $P^T$. The toric moduli space for the new model is identical to that of the original, under the exchange of $\mathcal{M}_t$ with $\mathcal{M}_c$. The discriminant ${\Delta}_c$ of the resulting model coincides precisely with ${\Delta}_t$ for the original, and vice versa.

Equivalently, as noted in \cite{Morrison:1995yh}, one can formulate the mirror model with twisted chiral charged fields coupled to twisted vector multiplets (with chiral field strength). The discussion of parameter spaces above is of course valid in this case as well, replacing chiral by twisted chiral (and vice versa) everywhere. The superconformal theories are in fact identical, and the prediction is that the partition functions must coincide exactly up to the ambiguity in their definition.

We will use this presentation for explicit computations. In other words, we will compare $Z_t$ for the model built from $P = \tilde T T$ with parameters $(q_a, c_{\tilde \imath})$ to $Z_t$ for the model built from $P^T = T^T \tilde T^T$ and parameters $(\tilde q_{\tilde a}, \tilde c_i)$ satisfying~\eqref{eq:mondivmap} but composed of twisted chiral charged fields, etc. The latter is equivalent to $Z_c$ for the model built from the same combinatorial data using chiral charged matter, etc. We will often abuse notation and refer to this as simply $Z_c$.

\subsection{Localization results}

To avoid excessive clutter, let us denote by $g = n - d$ and $\tilde g =
\tilde n - d$ the ranks of the gauge groups for the original and
the dual model, respectively. The $S^2$ partition function depending on twisted chiral parameters, $Z_t$, localizes to an integral of the classical action over the Coulomb branch, with integration measure provided by the $1$-loop determinants of quadratic fluctuations around this locus~\cite{Benini:2012ui, Doroud:2012xw}:
\be\label{eq:abelianpartition}
Z_t = \left(r \over r_0\right)^{c\over 3} \sum_{m^a \in \mathbb{Z} } \int \! {d^{g} \!\sigma \over \left(2\pi \right)^{g} } Z_{class}(\sigma, m) \prod_{i = 1}^n Z_i(\rho, \sigma, m),
\ee
where
\be
Z_{class} = \exp\left(-4\pi i \xi_a \sigma^a - i \theta_a m^a \right) = \prod_{a = 1}^g \left( q_a \right)^{i\sigma^a - {m^a \over 2}}\left( \bar q_a \right)^{i\sigma^a + {m^a \over 2}} .
\ee
and
\be
Z_i = {\Gamma\left( {\rho_i \over 2} - \sum_a Q_i^a \left( i \sigma^a +{1 \over 2} m^a \right) \right) \over \Gamma\left( 1- {\rho_i \over 2} + \sum_a Q_i^a \left( i \sigma^a - {1 
\over 2} m^a \right) 
\right) }.
\ee
The latter are the $1$-loop determinants of the matter multiplets around the Coloumb branch.

We will assume that a choice of $R$-charges with $\rho_i > 0$ has been made, which is always possible, and which implies that the integrand is non-singular over $\sigma \in \mathbb{R}^r$. The integrand of~\eqref{eq:abelianpartition} is meromorphic in each of the $\sigma^a$ variables and the integral can be evaluated by a multi-dimensional method of residues, 
where the contour---and thus which poles contribute---is chosen based on the values of the FI parameters. For example, when there is a single FI parameter, the contour can close in 
the upper half-plane if $|q|$ is sufficiently greater than $1$ and in the lower half-plane when $|q|$ is sufficiently less than $1$, corresponding to the Landau-Ginzburg and geometric phases, respectively. In the former case, only those poles at $i \sigma < 0$ contribute, while in the latter case only poles with $i \sigma > 0$ contribute.

If we make a different choice of $\rho_i$ by mixing with the gauge symmetry, $\rho_i \to \rho_i + \delta_a Q^a_i$, it is straightforward to see that the partition function changes only by an overall factor:\footnote{If the change in $\rho_i$ is such that $\rho_i >0$ no longer holds, then poles of the integrand will pass through the contour. We define the integral in this case by shifting the contour so that only the same poles contribute as when $\rho_i>0$.}
\be
Z_t \to \prod_{a =1}^g |q_a|^{\delta_a} Z_t.
\ee
As noted previously, this ambiguity in the partition function is an expected scheme-dependent effect. In particular, the choice of an $R$-symmetry is needed to define the coupling of the UV theory to the background metric on the sphere, and thus it is needed to define the regularization scheme. However, it has no effect on the scheme-independent quantities derived from the partition function.

The calculation of $Z_c$ for this model follows from exchanging chiral
fields for twisted chiral fields and twisted chiral fields strengths
for chiral field strengths, etc., and coupling this model to the same
sphere background. The localization for such a model was performed
in~\cite{Gomis:2012wy} for a Landau-Ginzburg theory and
in~\cite{Doroud:2013pka} for a gauge theory. Both results can be
summarized as an integral 
over the
Higgs branch, i.e. the space of orbits under the complexified gauge
group of the constant modes of
the twisted chiral fields. Note that for a Landau-Ginzburg theory, the
space of gauge orbits is the 
field space itself.  In particular,
\be\label{eq:doroudbmodel}
Z_c =\left( {r \over r_0}\right)^{c/3} \int\limits_{
{\mathbb{C}^{ n} / \left( \mathbb{C} ^\ast \right)^{
g}}} \!\!\!\!\!\! d\mathrm{vol}  ~ e^{\tilde W - \bar{\tilde
W}},
\ee
where $\tilde W$ is the superpotential~\eqref{eq:eW} but written in
terms of twisted chiral fields.
The measure on the space of gauge orbits, when the quotient is
nontrivial, follows from the flat measure on $\mathbb{C}^n$ after choosing a gauge slice via a finite-dimensional analog of the Fadeev-Popov procedure. 
To compare to the result of~\cite{Doroud:2013pka}, we partially fix the gauge with the standard $D$-term constraint:
\be
Z_c = \left( {r \over r_0}\right)^{c/3} \int d^{2n} \tilde \Phi~ \det \left( M^\dagger M \right) \prod_{a = 1}^g \delta\left( 2\mu^a - \xi^a \right)e^{\tilde W - \bar{\tilde
W}},
\ee
where
\be
\left(M^\dagger M \right)_{ab} = \sum_{i = 1}^n Q_i^aQ_i^b |\tilde\Phi_i|^2
\ee
is the Fadeev-Popov measure, and
\be
\mu^a = {1 \over 2} \sum_{i = 1}^n Q_i^a |\tilde \Phi_i|^2
\ee
is the $D$-term or moment map of the $a$-th $\mathbb{C}^\ast$ action. Additionally, we can restore the $R$-dependence to the twisted superpotential by scaling the fields
\be
\tilde \Phi_i \to \left( r \over r_0\right)^{\rho_i/2} \Phi_i,
\ee
recalling that under this scaling, the twisted superpotential has axial $R$-charge $2$, and also using~\eqref{eq:c3}. The result, apart from irrelevant numerical factors, is that of~\cite{Doroud:2013pka}:
\be
Z_c = \left( r \over r_0 \right)^{n - r} \int d^{2n} \tilde \Phi~ \det \left( M^\dagger M \right) \prod_{a = 1}^g \delta\left( 2\mu^a - \xi^a \right)e^{{ r \over r_0}\left(\tilde W - \bar{\tilde
W} \right)}.
\ee
The scale, $\mu$, appearing in~\eqref{eq:glsm} has been identified with $r_0^{-1}$.

There is no $\rho$-dependent ambiguity in this partition function analogous to that of $Z_t$. This is because a twisted chiral superfield is forced to have vanishing vector $R$-charge while the non-vanishing axial $R$-charge does not affect its coupling to the background metric~\cite{Closset:2014pda}.

However, the partition function does not respect the scaling symmetry
of the $c_{\tilde \imath}$,~\eqref{eq:rescale}. Instead, under
$c_{\tilde \imath} \to \lambda_i^{P_{\tilde \imath i}} c_{\tilde  \imath}$ 
the fact that $\tilde W$ is invariant if this is 
combined with~\eqref{eq:cstar}, $\Phi_i \to  \lambda_i^{-1}\Phi_i$,
shows that 
$Z_c$ transforms:
\be
Z_c \to\prod_{i = 1}^{n} \left| \lambda_i \right|^{-2} Z_c.
\ee
This, too, is an expected scheme-dependence. While the superpotential,
and therefore the IR fixed point, is invariant
under~\eqref{eq:rescale} and~\eqref{eq:cstar}, the UV GLSM is
not. Instead, this transformation acts a change of renormalization
scheme. The conclusion is that $Z_c$ as calculated from the UV depends
not on the invariant coordinates~\eqref{eq:eqt} but on the homogeneous
coordinates $c_{\tilde \imath}$.  As with the $\rho$ dependence of
$Z_t$, IR properties such as the Zamolodchikov metric depend only on
the invariant coordinates.

The scheme-dependence of $Z_t$ and that of $Z_c$ are of a different
character. One leads to dependence on the 
choice of $\rho_i$ and the other leads to dependence on the homogeneous coordinates of $\mathcal{M}_c$. Not surprisingly, then, $Z_t$ for a given theory will not be exactly equal to $Z_c$ of its mirror as given by~\eqref{eq:doroudbmodel}, since the former is explicitly a function of the invariant coordinates while the latter is independent of $\rho_i$. This is indeed what we find in the next section.

Before continuing, we remark that the sphere partition functions
$Z_{t/c}$ are insensitive to the splitting $P = \tilde T T$, and
therefore insensitive to any discrete gauge symmetries that result
from this splitting, apart from an overall numerical coefficient. This
can be argued in a couple of ways. First, the calculation of $Z_t$
proceeds through the Coloumb branch on which the action of these
discrete gauge factors is trivial. Alternatively, 
thinking of the partition function as the two-point function of the
identity operator, only untwisted sector states contribute.

Twisted sectors do contribute to the partition function on the orbifold of the sphere, or
equivalently on the sphere in the presence of defect
operators~\cite{Hosomichi:2015pia,Hosomichi:2017dbc} which can create
twisted sector states. These may be sensitive to the splitting $P =
\tilde T T$ and could act as a refined test of mirror
symmetry. Similarly, the elliptic genus may be sensitive to this
splitting~\cite{Benini:2013nda,Benini:2013xpa}. We will leave such
explorations to future work. Therefore, in the following, we will
restrict to models with only continuous gauge symmetries.  

\section{From $Z_t$ to $Z_c$}\label{sect:mirrorpartitions}
In the following, we will demonstrate that $Z_t$ of the model built from $P$ is equal, up to scheme-dependence, to that of $Z_c$ for the model built from $P^T$. Along the way, we will clarify somewhat the relationship between the latter and the Hori-Vafa mirror of the former, a relationship that has not seen much commentary.

To evaluate $Z_t$, we use the identity
\be
\int_0^\infty dt ~t^\mu J_\nu (t) = 2^\mu { \Gamma\left( {1 \over 2} \left( \mu + \nu + 1\right) \right) \over\Gamma\left( {1 \over 2} \left( -\mu  + \nu + 1\right) \right) }.
\ee
For our purposes, we restrict to $\nu \in \mathbb{Z}$. In that case, this identity holds when $-1 - | \nu | < \mathrm{Re}~\mu < {1 \over 2}$ with the first inequality required for 
convergence near 
$t = 0$ and the second required for convergence as $t \to \infty$. Setting
\be
\mu_i = \rho_i - 2i Q_i \cdot \sigma -1, \qquad \nu_i = - Q_i \cdot m, 
\ee
we can apply this identity if we restrict $0 < \rho_i < {3 \over 2}$. Furthermore, since $\nu_i \in \mathbb{Z}$, we can write the Bessel functions as
\be
J_\nu (t) = {1 \over 2\pi} \int_{-\pi}^{\pi} dy~ e^{i t \sin y - i \nu y}.
\ee
Changing variables $t_i = 2e^{x_i}$, we have
\bea
Z _t=&{1 \over \pi^n } \left( {r\over r_0} \right)^{c \over 3} \int \! \!d^nx \int_{-\pi}^{\pi} \! \! d^ny  \sum_{m^a} \int\! {d^g \! \sigma \over \left(2\pi\right)^g } \cr
&\exp\left(\rho_ix_i -2i \sigma \cdot \left( Q_i x_i + 2\pi \xi \right) + im \cdot \left(Q_i y_i - \theta \right) +\sum_{i=1}^n e^{x_i + iy_i} - e^{x_i - iy_i} \right).
\eea
In this form, as pointed out in~\cite{Benini:2012ui,Gomis:2012wy}, the partition function is that of the Hori-Vafa mirror~\cite{Hori:2000kt}. Specifically, following~\eqref{eq:doroudbmodel}, it is the sphere partition 
function of a twisted Landau-Ginzburg theory with fields $Y_i = x_i + iy_i$ and $\Sigma_a$ with twisted superpotential
\be
\tilde W_{HV} = -i \Sigma_a \left( Q_i^a Y_i - \log q^a \right) + \sum_{i=1}^n e^{Y_i}.
\ee
The imaginary part of a twisted chiral field strength multiplet, such
as $\Sigma_a$, is quantized.

The factors of
$e^{\rho_i x_i} = e^{{\rho_i \over 2} \left( Y_i + \bar Y_i \right)}$
can be thought of as modifying the measure, changing the variables in
terms of which this is flat from $Y_i$ to $e^{{\rho_i \over 2}Y_i}$.
The same change of variables was part of the prescription of Hori and
Vafa~\cite{Hori:2000kt} for calculating the periods of compact CICY
from an associated, non-compact toric CY.  These authors produced
indirect arguments for this change of variables, interpreted in flat
space as a selection of the universality class of the kinetic terms in
the dual model.

In the sphere partition function, they arise naturally.  This was
understood in~\cite{Gomis:2012wy,Benini:2012ui} as a consequence of
abelian duality on the sphere.  The sphere partition function depends
on the $R$ charges $\rho_i$ through a holomorphic dependence on
$\tilde m + i {\rho \over r}$ where $\tilde m$ are twisted masses for
the chiral matter fields.  The dual of a chiral field with twisted
mass was considered in~\cite{Hori:2000kt}, and indeed the effect is to
introduce a linear correction to the Hori-Vafa twisted superpotential 
$\delta \tilde W_{HV} = \tilde m Y$.  This superpotential correction,
however, vanishes in the flat space limit $r\to\infty$.

Another derivation of how these terms appear is described in appendix~\ref{sect:abelianduality}. Briefly, dualizing a chiral field with $R$-charge $\rho$ coupled to a background $R$-symmetry gauge field 
leads to a coupling in the dual theory of the form $\rho Y \epsilon^{\mu \nu} F^R_{\mu\nu}$. While there is no background $R$-symmetry gauge field on the sphere, the 
supersymmetrization of this term yields the desired linear twisted superpotential.

Continuing with $Z_t$, recall the integral form of the delta function
\be
\delta\left( x - y \right) = {1 \over 2\pi } \int dk ~e^{ik \left( x - y\right) },
\ee
where one can view this equality as occurring inside an integral for more rigor. Further, recall the Poisson summation formula
\be
\sum_{\lambda \in \Lambda} F(x + \lambda) = \sum_{\lambda^\ast \in \Lambda^\ast} \hat F( \lambda^\ast) {e^{2\pi i \lambda^\ast (x)} \over \mathrm{vol}(\Lambda^\ast) },
\ee
where $\Lambda$ is a lattice, $\Lambda^\ast$ is its dual, both of which are viewed as subsets of $\mathbb{R}^D$. Further,
\be
\hat F(k) := \int~d^Dx~e^{-2\pi i k(x)} F(x).
\ee
Using $F(x) = \delta(x)$, we have the periodic delta function
\be
\sum_{\lambda \in \Lambda} \delta(x + \lambda) = \sum_{\lambda^\ast \in \Lambda^\ast} {e^{2\pi i \lambda^\ast (x)} \over \mathrm{vol}(\Lambda^\ast) }.
\ee

Applying these formulae,
\bea
Z_t = {1 \over \pi^{d} } &\left( {r\over r_0} \right)^{c \over 3}  \sum_{m^a} \int d^n x \int_{-\pi}^\pi d^n y \cr
&~e^{ \left( x_i \rho_i + \sum_i e^{x_i + iy_i}- e^{x_i - iy_i} \right)} \prod_{a=1}^g\delta \left( Q^a_i x_i + 2\pi\xi^a \right)\delta\left( Q^a_i y_i - \theta^a + 2\pi m^a \right).
\eea
Given any solution, $a_i$ and $b_i$, to
\be\label{eq:variableshift}
\sum_{i=1}^n Q^a_i a_i = -2\pi \xi^a,  \qquad \sum_{i=1}^n Q^a_i b_i = \theta^a,
\ee
we can shift variables
\be
x_i \to x_i + a_i, \qquad y_i \to y_i + b_i,
\ee
in terms of which
\bea\label{eq:partitionHVform}
Z_t = {1 \over \pi^{d} } \left( {r\over r_0} \right)^{c \over 3} &\left(\prod_{i=1}^n |\tilde c_i|^{\rho_i} \right)\sum_{m^a} \int d^n x \int_{-\pi}^\pi d^n y \cr
&e^{ \left( x_i \rho_i + \sum_{i=1}^n \tilde c_ie^{x_i + iy_i}- \bar{\tilde c}_i e^{x_i - iy_i} \right)} \prod_{a=1}^g\delta \left( Q^a_i x_i \right)\delta\left( Q^a_i y_i + 2\pi m^a \right)
\eea
where $\tilde c_i = e^{a_i + i b_i}$. Note that our requirement on $a_i$ and $b_i$ translates to
\be
\prod_{i=1}^n \tilde c_i^{Q_i^a} = q^a,
\ee
which is the monomial-divisor mirror map~\eqref{eq:mondivmap}. Further, had we chosen a different solution to~\eqref{eq:variableshift}, $\tilde c'_i = \lambda_i \tilde c_i$, a shift of variables $ x_i + i y_i \to 
x_i + iy_i - \log 
\lambda_i$ would remove the dependence of $Z_t$ on $\lambda_i$. In other words, $Z_t$ doesn't depend on which choice of a solution to~\eqref{eq:variableshift} we use.

Furthermore, we can pause to comment on the the dependence of $Z_t$ on a choice of $R$-charges. Previously, for convergence, we stipulated that the $R$-charges lie in the 
range $0< \rho_i 
< {3 \over 2}$. However, now it can be seen that a shift of the $R$-charges by a linear combination of the gauge charges only multiplies $Z_t$ by powers of the FI parameters 
and does not 
affect convergence. Consider the shift $\rho_i \to \rho_i + \delta_a Q^a_i$. The delta function removes the dependence of the integrand on $\delta_a$. Additionally, the 
modification of the prefactor amounts to
\be
\prod_{i=1}^n |\tilde c_i|^{\rho_i+ \delta_aQ^a_i}  = \prod_{a=1}^g \left| q^a \right|^{\delta_a} \prod_{i=1}^n |\tilde c_i|^{\rho_i} ,
\ee
where the monomial divisor map was used. This is the same dependence on $\delta_a$ as was found from the definition of $Z_t$ in terms of residues.

The partition function in the form~\eqref{eq:partitionHVform} is still in the form of the Hori-Vafa mirror. In~\cite{Benini:2012ui,Gomis:2012wy} it was demonstrated that for Calabi-Yau hypersurfaces 
in $\mathbb{P}
^n$, the delta function constraints can be solved directly and the resulting partition function is that of an orbifold of a twisted LG theory with twisted superpotential given by the 
Greene--Plesser (or Batyrev--Borisov) dual to the 
original model, i.e., it is the dual from~\cite{Greene:1990ud} in the Landau-Ginzburg phase.

Another way to solve the delta function constraints will more clearly
relate the original model, and thus also its Hori-Vafa mirror, to the
combinatoric mirror of Batyrev and  Borisov. 
Recall that 
$Q_i^a$ span the kernel of the matrix $P$, therefore the delta functions enforce that $x \in \left(\mathrm{ker}P \right)^\perp \simeq \mathrm{im} P^T$. In turn, this implies 
there exists $\tilde x 
\in \mathbb{R}^{\tilde n}$ such that $x = P^T \tilde x$. However, $\tilde x$ is only determined up to $\mathrm{ker} P^T$, i.e. $\tilde x$ and $\tilde x + \tilde \delta_{\tilde a} \tilde 
Q^{\tilde a}$ 
yield the same $x$, where $\tilde Q^{\tilde a}$ span the kernel of $P^T$. Similar statements hold for $y$ modulo $2\pi \mathbb{Z}$. 

All of that is to say that we can write $Z_t$ as an integral over the $\tilde x$ and $\tilde y$ in $\mathbb{R}^{\tilde n} \times T^{\tilde n}$ modulo the action of the `gauge' symmetry: $\tilde x + i \tilde y \to \tilde x + i \tilde y + 
\left( \tilde 
\delta_{\tilde a} + i \tilde \gamma_{\tilde a} \right) \tilde Q^{\tilde a}$. The measure on this space follows from the flat measure on $\mathbb{R}^{\tilde n} \times T^{\tilde n}$ after fixing a gauge slice \'a la Fadeev-Popov.
\bea
Z_t = {1 \over \pi^{d} }\left( {r\over r_0} \right)^{c \over 3} &\left(\prod_{i=1}^n |\tilde c_i|^{\rho_i}\right) \int
\limits_{{\mathbb{R}^{\tilde n}\times T^{\tilde n}\over
    \mathbb{R}^{\tilde g}\times T^{\tilde g}} }\!\!\! d \mathrm{vol}(\tilde x, \tilde y)~e^{2\sum_{{\tilde{\imath}}} \tilde x_i } \cr
&\exp\left( \sum_{i=1}^n \left(\tilde 
c_i e^{\sum_{{\tilde{\imath}}}(\tilde x_{{\tilde{\imath}}} + i \tilde y_{{\tilde{\imath}}})P_{{\tilde{\imath}} i}} - \bar{\tilde c}_i e^{\sum_{{\tilde{\imath}}}(\tilde x_{{\tilde{\imath}}} - i \tilde 
y_{{\tilde{\imath}}})P_{{\tilde{\imath}} i}} \right)\right).
\eea
Here we have used $P \cdot \rho = (2, 2, \ldots, 2)^T$.  
The prefactor of $\exp\left(2\sum_{{\tilde{\imath}}} \tilde x_i
\right)$ 
can be incorporated into a change of the measure, which will now be
flat in terms of the variables
$\tilde \Phi_{{\tilde{\imath}}} = e^{\tilde x_{{\tilde{\imath}}} +
  i\tilde y_{{\tilde{\imath}}}}$,
on which 
the gauge action is via $\left(\mathbb{C}^\ast\right)^{\tilde g}$:
\be\label{eq:bmodelpartition}
Z_t = {1 \over \pi^{d} }\left( {r\over r_0} \right)^{c \over 3}  \prod_{i=1}^n |\tilde c_i|^{\rho_i} \!\!\! \int\limits_{ {\mathbb{C}^{\tilde n} / \left( \mathbb{C} ^\ast \right)^{\tilde g}}} \!\!\! 
d\mathrm{vol}  ~ \exp \left(\tilde W - \bar{\tilde W} \right),
\ee 
with
\be
\tilde W =  \sum_{i = 1}^{n} \tilde c_i \prod_{{\tilde{\imath}} = 1}^{\tilde n} \tilde \Phi_{{\tilde{\imath}}}^{P_{{\tilde{\imath}} i}}.
\ee
This is the superpotential of the combinatoric mirror to the original model.

Apart from irrelevant constants (that we have not been especially
careful to track and wchih can be absorbed into a rescaling of $r_0$),
~\eqref{eq:bmodelpartition} differs from~\eqref{eq:doroudbmodel} by
the prefactor $\prod_{i=1}^n |\tilde c_i|^{\rho_i}$. Owing to the
relationship $\sum_i P_{\tilde \imath i} \rho_i = 2$, this prefactor
is sufficient to remove the transformation of $Z_c$
under~\eqref{eq:rescale} (using the invariance of $\tilde W$ under
this combined with~\eqref{eq:cstar}). However, not surprisingly, it
introduces the same $\rho$-dependent scheme-dependence 
exhibited by 
$Z_t$.
Since the disagreement
between~\eqref{eq:bmodelpartition} and~\eqref{eq:doroudbmodel} is
precisely of the expected form, we conclude that $Z_{t/c}$ is
consistent with the mirror symmetry conjecture. 

We have made our basic calculation from first principles, using localization, but
it would also be interesting to know if our computation could, in the alternative,
be based on methods of Givental \cite{MR1403947,complete1}, who also used localization to
obtain his basic results.

\subsection*{Acknowledgements}

It is a pleasure to thank
Ron Donagi,
Jaume Gomis,
Zohar Komargodski,
Mauricio Romo,
Nati Seiberg,
and
Eric Sharpe
for discussions and correspondence.
DRM and MRP thank the Institut Henri Poincar\'e for hospitality during
the early stages of this project; in addition,
MRP thanks
UC Santa Barbara, the CERN Theory group, and the HET group at the
Weizmann Institute of Science, and 
DRM thanks the Kavli Institute for 
Theoretical Physics and the Mathematical Sciences Research Institute
for hospitality during other stages of the project.
The work of DRM was supported by the Centre National de la Recherche Scientifique
(France), by NSF grants PHY-1307514, 
DMS-1440140,
PHY-1620842, and 
PHY-1748958
(USA), and by a Fellowship in Theoretical Physics from the Simons Foundation 
[award \#562580].
The work of MRP was supported by NSF grant PHY-1521053.
Any opinions, findings, and conclusions or
recommendations expressed in this material are those of the authors
and do not necessarily reflect the views of the National Science
Foundation. 

\appendix

\section{Demonstration That $Z_c$ Solves the $A$-system}\label{sect:asystem}
In~\cite{Halverson:2013eua}, it is shown that $Z_t$ satisfies a set of differential equations in the parameters, the \emph{$A$-system}. This equation is most compactly written in terms of an auxiliary 
function defined 
as
\be
\Psi_t = \left(\prod_{i = 1}^n \left|\tilde c_i \right|^{-\rho_i} \right)Z_t(q^a, \bar q^a).
\ee
The set of equations is
\bea\label{eq:asystem}
\prod_{\{i | Q_i^a > 0\} }\left( {\partial \over \partial \tilde c_i }\right)^{Q_i^a} \Psi_t &= \prod_{\{i | Q_i^a < 0\} } \left({\partial \over \partial \tilde c_i }\right)^{|Q_i^a|} \Psi_t, \quad \forall a, \cr
\sum_{i = 1}^{ n} P_{\tilde \imath i } \tilde c_i {\partial \over \partial \tilde c_i} \Psi _t &= -\Psi_t, \quad \forall \tilde \imath \ .
\eea
The function $\Psi_t$, we have demonstrated, is precisely $Z_c$,~\eqref{eq:doroudbmodel}, of the mirror theory. In this presentation, it is quite straightforward to see that $Z_c$ solves the $A$-system, and below we will give the details for completeness.
Ignoring numerical factors,
\be
\Psi_t = Z_c( \tilde c_i, \bar{\tilde c}_i) = \int\limits_{ {\mathbb{C}^{\tilde n} / \left( \mathbb{C} ^\ast \right)^{\tilde g}}} \!\!\! 
d\mathrm{vol}  ~ \exp \left(\tilde W - \bar{\tilde W} \right),
\ee 
with
\be
\tilde W =  \sum_{i = 1}^{n} \tilde c_i \prod_{{\tilde{\imath}} = 1}^{\tilde n} \tilde \Phi_{{\tilde{\imath}}}^{P_{{\tilde{\imath}} i}}.
\ee
The only $\tilde c_i$ dependence of $\Psi_t$ is from $\tilde W$.
\be
{\partial \over \partial \tilde c_i} \tilde W = \prod_{{\tilde{\imath}} = 1}^{\tilde n} \tilde \Phi_{{\tilde{\imath}}}^{P_{{\tilde{\imath}} i}}.
\ee
Therefore,
\be
\left({\partial \over \partial \tilde c_i}\right)^{Q_i^a} e^{\tilde W} = e^{\tilde W}\prod_{{\tilde{\imath}} = 1}^{\tilde n} \tilde \Phi_{{\tilde{\imath}}}^{P_{{\tilde{\imath}} i} Q_i^a}.
\ee
And,
\bea
\prod_{\{i | Q_i^a > 0\} }\left( {\partial \over \partial \tilde c_i}\right)^{Q_i^a} e^{\tilde W} &= e^{\tilde W} \prod_{{\tilde{\imath}} = 1}^{\tilde n}\left( \tilde \Phi_{{\tilde{\imath}}} 
\right)^{ \sum\limits_{ \{i 
| Q_i^a > 0 \}}P_{{\tilde{\imath}} i} Q_i^a} \cr
& = e^{\tilde W} \prod_{{\tilde{\imath}} = 1}^{\tilde n}\left( \tilde \Phi_{{\tilde{\imath}}} \right)^{ -\!\!\! \sum\limits_{\{i | Q_i^a < 0 \}}P_{{\tilde{\imath}} i} Q_i^a} \cr
& = \prod_{\{i | Q_i^a < 0 \}}\left( {\partial \over \partial \tilde c_i}\right)^{|Q_i^a|} e^{\tilde W}, 
\eea
where we've used $\sum_{i} P_{{\tilde{\imath}} i} Q_i^a=0$. This is sufficient to show that the first of~\eqref{eq:asystem} holds.

To show the second holds, we observe
\be
\sum_{i = 1}^{ n} P_{\tilde \imath i } \tilde c_i {\partial \over \partial \tilde c_i} \Psi _t~ = \int\limits_{ {\mathbb{C}^{\tilde n} / \left( \mathbb{C} ^\ast \right)^{\tilde g}}} \!\!\! 
d\mathrm{vol} ~ \tilde \Phi_{\tilde \imath} {\partial \over \partial \tilde \Phi_{\tilde \imath}} \exp \left(\tilde W - \bar{\tilde W} \right), \quad \forall \tilde \imath.
\ee 
After integrating by parts, the result follows.

\section{Abelian Duality on and off the Sphere}\label{sect:abelianduality}
In this appendix, we argue for the existence on the sphere of the linear twisted superpotential $\tilde W \sim \rho Y$ for fields $Y$ dual to chiral fields with $R$-charge $\rho$. As stated in the body, this argument has appeared previously in multiple forms. The following is a slight modification of the argument appearing in~\cite{Gu:2018fpm}.

Consider a two-dimensional theory of a complex scalar field:
\be
S = \int d\phi \wedge \ast d\bar \phi = \int  d\sigma \wedge \ast d\sigma + \sigma^2 d\theta \wedge \ast d\theta.
\ee
The change of variables from the first equation to the second, $\phi = \sigma e^{i\theta}$ will be very badly behaved around $\phi = 0$. Nevertheless, away from this point, we can 
classically dualize 
the $U(1)$ isometry under which $\theta \to \theta + \epsilon$. To do so, consider instead the following action for a $1$-form $c$ and a Lagrange multipler $\lambda$, 
ignoring the kinetic 
action for the $\sigma$ field which plays no part:
\be
S = \int \sigma^2 c \wedge \ast c - 2c \wedge d \lambda.
\ee
Integrating out $\lambda$ implies $c$ is closed and therefore exact (in $\mathbb{R}^2$ for now and later in $S^2$ also). Therefore, we reproduce the original action. Instead, 
integrating out 
$c$ yields
\be
\ast c = {1 \over \sigma^2} d\lambda \quad \Rightarrow S = -\int {1 \over \sigma^2} d\lambda \wedge \ast d\lambda.
\ee

We can repeat this calculation with the current associated to $\theta \to \theta + \epsilon$ coupled to a background gauge field $A$. Our starting point is
\be
S = \int \sigma^2  \left( c - A \right) \wedge \ast \left( c - A\right) - 2 c \wedge d\lambda.
\ee
Integrating out $c$ yields
\be\label{eq:bosonictdualwithA}
S =- \int {1 \over \sigma^2} d\lambda \wedge \ast d\lambda + 2\lambda dA.
\ee

The supersymmetrization of this starts with a chiral superfield $\Phi$ with canonical K\"ahler potential $|\Phi|^2$. Away from $\Phi = 0$, we may define $\Phi = e^\Pi$, where 
$\Pi$ is also 
chiral. To dualize the phase of $\Pi$, analogous to $\theta$ above, we replace $\Pi$ with an unconstrained, real superfield and add a Lagrange multiplier to reinstate the chiral 
constraint:
\be
\mathcal{L} = \int d^4\theta  ~e^{\Pi + \bar \Pi } \to \int d^4 \theta~ e^{2B} - 2B \left( Y + \bar Y\right).
\ee
Integrating out $Y$ ensures that $B = \Pi + \bar \Pi$. Instead, integrating out $B$ we find
\be
\mathcal{L} = -\int d^4 \theta ~ \left( Y+ \bar Y \right) \log \left( Y + \bar Y \right).
\ee
If $\Pi$ is coupled to a background vector supermultiplet, the appropriate supersymmetrization of the coupling of a global, flavor symmetry to a background field, then the 
above is modified to
\bea
\mathcal{L} &= -\int d^4 \theta ~ \left( Y+ \bar Y \right) \log \left( Y + \bar Y \right) - 2 \left(Y + \bar Y \right) V \cr
& = -\int d^4 \theta ~ \left( Y+ \bar Y \right) \log \left( Y + \bar Y \right) - 2\int d^2 \tilde \theta Y \Sigma ~+ ~c.c.
\eea
However, if $\Phi$ has $R$-charge $\rho$ and is instead coupled to a background $R$-symmetry gauge field, we still expect a coupling of the dual $Y$ to the field strength of 
this gauge field 
from~\eqref{eq:bosonictdualwithA}, but the supersymmetrization of this coupling will not be $Y\Sigma$ because the $R$-symmetry current is not contained in an ordinary 
linear multiplet. It is 
contained in the $R$-multiplet, and so the corresponding gauge field is contained in the gravity multiplet.

The supersymmetric coupling responsible is
\be
\rho \int d^2 \tilde \theta~ \tilde{\mathcal{E}} \tilde{\mathcal{R}} Y ~+ ~ c.c.,
\ee
where $\tilde{\mathcal{E}}$ is the twisted supersymmetric density and $\tilde{\mathcal{R}}$ is a twisted chiral curvature superfield containing the Ricci scalar and the 
curvature of the $U(1)_V$ 
gauge field, among other terms. When evaluated in the supersymmetric sphere background, this coupling gives precisely the linear twisted superpotential that effects the change of fundamental variable in agreement with the 
Hori-Vafa 
prescription. More details about this coupling can be found in~\cite[eq. (6.73)]{Closset:2014pda} and~\cite[eq. (3.31)]{Gerchkovitz:2014gta}. One salient feature to note is that this coupling does not survive the flat-space limit, and so it is not present in the original~\cite{Hori:2000kt}.

\newpage
\bibliographystyle{utphys}
\bibliography{bib}
\end{document}